\pdfoutput=1
\documentclass[sigconf]{acmart}

\AtBeginDocument{%
  \providecommand\BibTeX{{%
    \normalfont B\kern-0.5em{\scshape i\kern-0.25em b}\kern-0.8em\TeX}}}

\copyrightyear{2021}
\acmYear{2021}
\setcopyright{acmcopyright}

\acmConference[ESEM '21]{ACM / IEEE International
Symposium on Empirical Software Engineering and Measurement (ESEM)}{October
11--15, 2021}{Bari, Italy}
\acmBooktitle{ACM / IEEE International Symposium on Empirical Software Engineering
and Measurement (ESEM) (ESEM '21), October 11--15, 2021, Bari, Italy}
\acmPrice{15.00}
\acmDOI{10.1145/3475716.3475787}
\acmISBN{978-1-4503-8665-4/21/10}

\usepackage[normalem]{ulem} 
\usepackage{graphics} 
\usepackage{fancybox}
\usepackage{multirow}
\usepackage{flushend}
\usepackage{booktabs}
\usepackage{tabularx}
\usepackage{comment}
\usepackage{array}
\usepackage[flushleft]{threeparttable}
\usepackage{mdframed}
\graphicspath{{}{images/}{dia/}}
\DeclareGraphicsExtensions{.pdf,.png}
\usepackage{courier}
\usepackage{hyperref}
\usepackage{color}
\usepackage{xcolor}
\usepackage{alltt}
\usepackage{subfig}
\usepackage{graphicx}
\usepackage{float}
\usepackage{verbatim}
\usepackage{mathtools}
\usepackage{url}
\usepackage{paralist}
\usepackage{soul}
\usepackage{balance}
\usepackage[outercaption]{sidecap}    
\usepackage{enumitem}
\usepackage{pifont}
\usepackage[many]{tcolorbox}

\newcounter{o}
\def\bf{\textbf}

\def\fig {Figure~}

\def\tbl {Table~}

\def\sec {Section~}

\def\it{\textit}

\newcommand{\nd}{\vspace{1mm}\noindent}
\newcommand{\urls}[1]{{\scriptsize\url{#1}}}
\newcommand{\emt}[1]{\emph{``#1''}}

\newtcolorbox
{mybox}[2][]{colbacktitle=red!10!white,
colback=blue!10!white,coltitle=black!70!black,
title={#2},fonttitle=\bfseries,#1}

\setcopyright{none}
\renewcommand\footnotetextcopyrightpermission[1]{}
\setcopyright{none}
\makeatletter
\renewcommand\@formatdoi[1]{\ignorespaces}
\makeatother

\begin{document}

\title[A Survey-Based Qualitative Study to Characterize Expectations of Software Developers from Five Stakeholders]{A Survey-Based Qualitative Study to Characterize Expectations of Software Developers from Five Stakeholders}

\author{Khalid Hasan$^*$, Partho Chakraborty$^*$, Rifat Shahriyar$^*$, Anindya Iqbal$^*$, Gias Uddin$^{\mathsection}$}
\affiliation{%
\institution{Bangladesh University of Engineering and Technology, Dhaka, Bangladesh$^*$\\
University of Calgary, Alberta, Canada$^{\mathsection}$}
\country{}
}
\email{72.khalidhasan@gmail.com, shuvopartho@gmail.com, rifat@cse.buet.ac.bd, anindya@cse.buet.ac.bd, gias.uddin@ucalgary.ca}

\renewcommand{\shortauthors}{Khalid et al.}

\begin{abstract}
% \gias{ ESEM you need a structured abstract. See \urls{https://conf.researchr.org/track/esem-2021/esem-2021-technical-papers#How-to-Submit}. An example of a structured abstract is this: \urls{https://giasuddin.files.wordpress.com/2020/12/esem2020a-on-the-use-of-c-unsafe-code-context-an-empirical-study-of-stack-overflow.pdf}
% }
\textbf{\it{Background}.} Studies on developer productivity and well-being find that the perceptions of productivity in a software team can be a socio-technical problem. Intuitively, problems and challenges can be better handled by managing expectations in software teams. %It is, however, not known the expectations of developers in software teams.  
% In software development teams, expectations can regulate the success or failure of a project. Assuring that the different stakeholders (e.g., managers, developers, etc.) have the same expectations, open lines of communication can thus help permit a successful project. A study on the multi-faceted expectations in software teams can reveal the perceptions and scopes of improvement of similar types.
\textbf{\it{Aim.}} Our goal is to understand whether the expectations of software developers vary towards diverse stakeholders in software teams. 
\textbf{\it{Method.}}  We surveyed 181 professional software developers to understand their expectations from five different stakeholders: \begin{inparaenum} 
\item organizations, 
\item managers, 
\item peers, 
\item new hires, and 
\item government and educational institutions.
\end{inparaenum} The five stakeholders are determined by conducting semi-formal interviews of software developers. We ask open-ended survey questions and analyze the responses using  open coding.
% We aim to understand the expectations among software development teams from two surveys to bring a balance in the software industry. To understand the expectations in software teams, we conducted two surveys involving 181 software correspondents. By using an open coding approach and manual checks, we have categorized responses into key expectations.
\textbf{\it{Results.}} We observed 18 multi-faceted expectations types. While some expectations are more specific to a stakeholder, other expectations are cross-cutting. For example, developers  expect work-benefits from their organizations, but expect the adoption of standard software engineering (SE) practices from their organizations, peers, and new hires. 
% Developers also expect their governments to regulate policy to improve their job stability and from educational institutions to adopt industry-oriented teaching. 
% From the response of getting the expectations from an organization, we have observed that workplace culture needs to be dominantly focused on as per 48.78\% of respondents followed by proper management and work benefits. Again, most employees want their managers to have people related skills and manage their tasks properly. Moreover, 55.29\% employees expect their peers to be supportive, followed by the second most popular category, sincerity. On top of that, the new hires are expected to focus mostly on their learning by 52.94\% of participants. The second most popular category by 47.06\% of the respondents in the posts is used for 'Knack in software engineering (SE) practice'. Also, they are expected to be sincere. Finally, the major expectation from the government is job-related 'policies and facilities' by 63.08\% of participants. Also, participants want their government to be focused on the software industry and job availability. 77.14\% of the respondents expect industry oriented teaching from the universities.
\textbf{\it{Conclusion.}} Out of the 18 categories, three categories are related to career growth. This observation supports previous research that happiness cannot be assured by simply offering more money or a promotion. Among the most number of responses, we find expectations from educational institutions to offer relevant teaching and from governments to improve job stability, which indicate the increasingly important roles of these organizations to help software developers. This observation can be especially true during the COVID-19 pandemic. 
%(which is when our surveys were conducted).  

\end{abstract}




\maketitle

\section{Introduction}\label{sec:introduction}
Software development is a complicated task that requires critical
thinking, deep technical and domain-specific background and close collaboration
and communication with team members.
Given the importance of software in our every day life, from national policy
making to our personal life, ensuring the productivity and well-being of
software developers are important so that quality software is built properly~\cite{Albrecht-MeasureApplicatonDevelopmentProductivity-IBM1979,Paiva-FactorsProductivitySoftDev-CSSE2010}. At the same time, the retention period of a software developer in a company is generally low compared to other industries. While such developer turnover could be due to many factors, their turnover could also be due to  lack of support to manage their expectations and happiness in software teams~\cite{Graziotin-HappinessProductivitySoftwareEngineer-BookRethinkingProductivity2019}.

% On the other hand, the retention period of developers in a company is significantly low compared to other industries and the almost every departure costs immensely to the company. The problem is more pronounced in developing countries where, apart from switching to another company, a large number of developers migrate to developed countries in a regular manner. Hence, this is very important to have clear mutual understanding among the developers and the management/owners that would create better working environment. Eventually, it is likely to have significant impact on the retention of developers. 
A significant body of research has devoted to understand the factors that influence developers' 
productivity in software teams~\cite{Meyer-CharacterizeSoftDevPerceptionOfProductivity-ESEM2017,Nguyen-AnalysisTrendsProductivityCost-PMSE2011,Albrecht-MeasureApplicatonDevelopmentProductivity-IBM1979,Meyer-SoftDevPerceptionProductivity-FSE2014,Paiva-FactorsProductivitySoftDev-CSSE2010,Ko-WhyWeShouldNotMeasureProductivity-BookRethinkingProductivity2019,Perry-PeopleOrganizationProcessImprovement-IEEESw1994,Baruch-SelfPerformanceDirectManager-ManagerialPsychology1996,Chong-InterruptionsOnSoftwareTeams-CSCW2006,Czerwinski-DiaryStudyTaskSwitchingInterruptions-CHI2006,Parnin-CueForResumingInterruptedProgrammingTasks-CHI2010}. 
Recent such studies are mostly conducted at large organizations like Microsoft by 
focusing on multiple scenarios such as listening to music while working, being interrupted while at work, etc. 
The well-being of developers is studied within the context of productivity and 
socio-technical needs~\cite{Meyer-SoftDevPerceptionProductivity-FSE2014} and COVID-19~\cite{ralph2020pandemic}, which show that 
developers suffer, like everyone else, to meet expectations or 
to be productive during stressful moments or situations with interruptions~\cite{Bailey-EffectsOfInterruptionsPerformance-Interact2001,Chong-InterruptionsOnSoftwareTeams-CSCW2006}.   

The related studies on developer productivity and well-being offer valuable information 
to improve the productivity and well-being of developers by improving specific situations (e.g., by reducing interruptions). 
However, if we take a step back and think of this from a high-level perspective, we could also benefit from insights like 
overall expectations of developers in software teams from the major stakeholders. Such stakeholders can be managers, peers, the organization 
itself, or even the governments. Given such high-level perspective, it is then intuitive to frame this query from the viewpoint of understanding 
the expectations of developers from the diverse stakeholders that the developers perceive as important to make progress in their work and career.
Indeed, as Graziotin and Fagerholm~\cite{Graziotin-HappinessProductivitySoftwareEngineer-BookRethinkingProductivity2019} argue, the happiness 
of developers is paramount to improve their productivity in software teams. 

In this paper, we attempt to understand the expectations of software developers in software teams and 
whether we can produce a catalog of those expectations. Given this is a wide research question that is impossible to properly answer within 
the scope of a single research paper, we focus on learning the expectations from five different stakeholders: \begin{inparaenum} 
\item organizations, 
\item managers, 
\item peers, 
\item new hires, and 
\item government and educational institutions.
\end{inparaenum} The five stakeholders are determined by conducting literature review (see \sec\ref{sec:related-work}) 
and by conducting semi-formal interviews of software developers in software teams (see \sec\ref{study_setup}). Our goal is to 
learn about the expectations from professional software developers. We surveyed total 181 developers from companies of all types (i.e, small, medium, and large).
We analyze the survey responses using card sorting and open coding.   

We observed multi-faceted expectations among developers (see \sec\ref{sec:results}), 
which we group into 18 expectation types. 
While some expectations are more specific to a stakeholder, other expectations 
are cross-cutting across the stakeholders. 
For example, developers specifically expect `work-benefits' from their
organizations, but expect the adoption of standard software engineering (SE)
practices from their organizations, peers, and new hires. Developers also expect
their governments to regulate policy to improve their job stability and from
educational institutions to adopt industry-oriented teaching.
The 18 expectation types are groped under six themes. 
Four themes belong to work-related expectations: \begin{inparaenum}
\item Well-being (e.g, work benefits to support work-life balance),
\item Leadership (e.g., policy support from the government and tech savviness from team leads),
\item Practice (e.g., adoption of standard software engineering (SE) practices by all team members), and 
\item Productivity (e.g., proper measures within the organization to support goal achievement).  
\end{inparaenum} Two themes belong to the career-related expectations: \begin{inparaenum}
\item Career growth (e.g., opportunities for moving up the ladder), and
\item Education (e.g., industry-focused training in the education institutes).
\end{inparaenum}
% 
% We find that workplace culture on as per 48.8\% 
% of respondents followed by proper management and work benefits. 
% Most employees want their managers to have people related skills and manage their tasks properly. 
% Moreover, 55.3\% employees expect their peers to be supportive, 
% followed by the second most popular category, sincerity. On top of that, 
% the new hires are expected to focus mostly on their learning by 52.9\% of participants. 
% The second most popular category by 47.1\% of the respondents in the posts is used for 
% 'Knack in software engineering (SE) practice'. 
% Finally, the major expectation from the government is job-related 'policies and facilities' 
% by 63.1\% of participants.  
% 77.1\% of the respondents expect industry oriented teaching from the universities.

We find that out of the 18 categories, three categories are related to career
growth. This observation supports previous research that happiness cannot be
assured by simply offering more money or a promotion. Among the most number of
responses, we find expectations from educational institutions to offer relevant
teaching and from governments to improve job stability, which indicate the
increasingly important roles of these organizations to help software developers.
This observation can be especially true during the COVID-19 pandemic, which is when our study was conducted.
 
\nd\bf{Replication Package.} \url{https://cutt.ly/MbBy0aU}

\section{Related Work}\label{sec:related-work}
Related work can broadly be divided into two categories: \begin{inparaenum}
Developer/team  \item productivity (\sec\ref{sec:studyDevProductivity}), and
\item well-being (\sec\ref{sec:studyDevwell-being}).
%\item Challenges faced by developers in software teams. 
%\item Expectation management of developers.
\end{inparaenum} %We discuss selected related work below.
\subsection{Research on Developer Productivity}\label{sec:studyDevProductivity}
Significant research efforts have devoted to understand the factors affecting developers' productivity and 
the means and techniques that can be used to improve productivity of software developers~\cite{Meyer-CharacterizeSoftDevPerceptionOfProductivity-ESEM2017,Nguyen-AnalysisTrendsProductivityCost-PMSE2011,Albrecht-MeasureApplicatonDevelopmentProductivity-IBM1979,Meyer-SoftDevPerceptionProductivity-FSE2014,Paiva-FactorsProductivitySoftDev-CSSE2010,Ko-WhyWeShouldNotMeasureProductivity-BookRethinkingProductivity2019,Perry-PeopleOrganizationProcessImprovement-IEEESw1994,Baruch-SelfPerformanceDirectManager-ManagerialPsychology1996,Chong-InterruptionsOnSoftwareTeams-CSCW2006,Czerwinski-DiaryStudyTaskSwitchingInterruptions-CHI2006,Parnin-CueForResumingInterruptedProgrammingTasks-CHI2010,Nguyen-AnalysisTrendsProductivityCost-PMSE2011}.  
The assessment of productivity is a multi-faceted problem, as the originally observed by 
Albrecht~\cite{Albrecht-MeasureApplicatonDevelopmentProductivity-IBM1979} who formulated a productivity measure at IBM based on several key variables to projects, 
such project size, requirements, etc. They find that a disciplined process was necessary to increase productivity. 
Paiva et al.~\cite{Paiva-FactorsProductivitySoftDev-CSSE2010} identified 35 influence 
factors on developers' productivity, such as capability and experience, etc. 
% Nguyen et al.~\cite{Nguyen-AnalysisTrendsProductivityCost-PMSE2011} find that 
% average software productivity has increased by six times over the last 40 years. 

Recently, Sadowski et al.~\cite{Sadowski-SoftDevProductivityFramework-RethingDevProd2019} proposed a software development productivity framework by 
focusing on three dimensions: velocity (how fast work gets done), quality (how well work gets done), and satisfying (how well it was perceived).   
Indeed, to improve developers' productivity, we first need to understand how they perceive of their productivity. 
Previously, Meyer et al.~\cite{Meyer-SoftDevPerceptionProductivity-FSE2014} found that developers perceive their days as productive when they complete many or 
big tasks without significant interruptions or context switches. However, 
the observational study showed that developers still performed significant and big tasks while being interrupted. In general, 
interruptions can be disruptive to the productivity of software developers, especially while they work alone instead of in a 
team (e.g., pair programming)~\cite{Chong-InterruptionsOnSoftwareTeams-CSCW2006}.
Czerwinski et al.~\cite{Czerwinski-DiaryStudyTaskSwitchingInterruptions-CHI2006} report on a diary study of the activities of software developers to understand 
and characterize how people interleave multiple tasks amidst interruptions. They find that task complexity, task duration, length of absence, number of interruptions, and 
task type influence the perceived difficulty of switching back to tasks. Parnin and DeLine~\cite{Parnin-CueForResumingInterruptedProgrammingTasks-CHI2010} look for cues on 
how developers resume work after they are interrupted. 
% They surveyed 371 
% programmers on their nature of tasks, interruptions, task suspension, and resumption strategies.
They find that developers rely on heavy note-taking across 
several types of media to help them resume their interrupted work. 

Meyer et al.~\cite{Meyer-CharacterizeSoftDevPerceptionOfProductivity-ESEM2017,Meyer-DesinRecommendationsForSelfMonitoringSoftwareWorkplace-CHI2017} 
characterized the perceptions of productivity at Microsoft. 
In surveys of 413 developers at Microsoft,  they identified six groups of developers with similar perceptions of productivity: social, lone, focused, balanced, leading, and goal-oriented 
developers. Based on the characterization, the authors argue for personalized recommendation system for developers to assist in their productivity improvement.   
In parallel, they analyzed the impact of self-monitoring to improve the productivity of knowledge workers by inferring design elements for workplace self-monitoring. 
They implemented a technology called WorkAnalytics. Based on the deployment of the technology, 
they argue for diverse metrics and actionable insights to 
support self-monitoring. In this vein, a preliminary cognitive support framework based on bots was 
discussed by Storey and Zagalsky~\cite{Storey-DisruptDevProductivityBot-FSE2016}. Constant monitoring of interruptions and 
offering support via simple dimming of LED lights to reduce the interruption were also found to increase the productivity of the developers~\cite{Zuger-ReduceInterruptionsFlowlight-RethingDevProd2019}. Like other professions, the COVID-19 pandemic has also negatively affected the productivity of software developers. Ralph et al.~\cite{ralph2020pandemic} recommended 
organizations to not rely on traditional measures of software productivity during COVID-19. This recommendation is timely, because as 
Baruch~\cite{Baruch-SelfPerformanceDirectManager-ManagerialPsychology1996} finds, there is a correlation between an employee's self-appraisal and his/her direct manager's 
appraisal. As such, organizations may need to instead ask the developers to self-judge their productivity during the COVID-19. Indeed,     
Ko~\cite{Ko-WhyWeShouldNotMeasureProductivity-BookRethinkingProductivity2019} urged not to quantify productivity because such quantification can 
warp incentives if not measured well or can influence sloppy management.

Our work takes motivations from the above work that the productivity of developers can depend on how they manage their expectations in teams. 
We produce a catalog of expectation types by surveying professional software developers, which may support better productivity and well-being of developers.

\subsection{Research on Developer Well-being}\label{sec:studyDevwell-being}
Research on the well-being of software developers and the factors 
affecting developers' well-being is limited. In general, developers' are not happy when they are interrupted~\cite{Meyer-SoftDevPerceptionProductivity-FSE2014,Bailey-EffectsOfInterruptionsPerformance-Interact2001}. Perry et al.~\cite{Perry-PeopleOrganizationProcessImprovement-IEEESw1994} show that software 
developers spend significant time with their coworkers on non-coding activities, such as chats and in-person communication. They 
also find that the reluctance of developers to use emails can impact the development processes. Chong and Siino~\cite{Chong-InterruptionsOnSoftwareTeams-CSCW2006} find that developers are more resilient to interruptions while working in pair. With the recent advances in 
technology, pair programming is entirely possible while working remotely. Therefore, software developers may adopt more pair programming to improve their well-being and happiness while 
working from home. Developers' well-being can be better supported by making them more happy, which in turn is necessary to make them productive. 
Indeed, Graziotin and Fagerholm~\cite{Graziotin-HappinessProductivitySoftwareEngineer-BookRethinkingProductivity2019} argue that making software developers happy is very important to 
improve their productivity. 
% Perry et al.~\cite{Perry-PeopleOrganizationProcessImprovement-IEEESw1994} conducted two experiments to discover how software 
% developers spend their time. They find that non-coding activities, such as chats and in-person communication can use up significant development time. They 
% also find that the reluctance of developers to use emails can impact the development processes. 

%\subsection{Happiness and Joys of Software Developers}
% related work on productivity and happiness of software developers 
Sentiment analysis of various SE artifacts (e.g., code review, developer forums, JIRA issues) is found to be effective to learn about the success of a code reviews, to give developers better insights, and to support development activities~\cite{Ikram-SentimentCodeReview-IST2018,Uddin-OpinionValue-TSE2019,Uddin-SurveyOpinion-TSE2019}. Indeed, emotions experienced and expressed by software developers can offer reliable indication of their productivity and other physical or mental states while working in a software team. In their study, Graziotin et al.~\cite{Graziotin2018} identified the causal relationship between (un)happiness and productivity. They found that unhappiness takes its toll on low focus, inadequate performance, reduced skills, fatigue, and decision-making problems. Consequences like this hit directly at the core of the software development activity, as it is inherently intellectual.
To find the answer to the question `Are Happy Developers More Productive?'  Graziotin et al.~\cite{Graziotin2013} conducted a study to identify the relation of affective states (emotions and moods) of software developers and their productivity. They found that happy developers reported that they are productive more often than unhappy developers. In another article~\cite{Graziotin2014}, they suggested some steps to improve the affects of developers. One of these suggestions is, managers should respect the uniqueness of their subordinates. 

In this paper, we focus on learning the multi-faceted expectations of developers in software teams. Intuitively, a better managed team can support better expectation management. Challenges in development environment can be context-oriented (e.g., problems using a software library~\cite{Uddin-HowAPIDocumentationFails-IEEESW2015} vs. problems using a new programming language~\cite{Partha-NewLangSO-IST2021}, etc.). We report the overarching expectations in software teams irrespective of specific development contexts. Such insights can help the formulation of specific goals across teams.

%\rifat{The current related work section looks great.}

%\subsection{Research on Challenges faced by software developers}
%related work on software development challenges

% \subsection{Expectation management in software development}
% any related work that specifically focused on understanding software development expectations, look at works on career aspirations as well.

\section{Study Setup}
\label{study_setup}
We followed four major steps in our study: \begin{inparaenum}
\item We conduct a series of semi-formal interviews with software developers in industry to understand the types of expectations 
they have towards different stakeholders that influence their daily development activities.
\item We formulate a set of questions to based on the interview data and by analyzing how the questions could complement existing literature.
\item We conduct a survey with another set of industrial software developers different from the interview participants.
\item We analyze and report the survey responses.
\end{inparaenum} We describe the steps below. 

\subsection{Interviews 
\& Survey Question Formulation}
The goal of
the interview session was to prepare the survey questions. Eight individual participants from four software companies were
interviewed. First, we designed an initial list of survey questions in Google form by consulting previous studies that focused on software development practices in various countries like 
Canada, Turkey, Netherlands, New Zealand, etc.~\cite{Garousi2013,Garousi2015,Vonken2012,Wang2018}. We specifically focused on these studies 
because they investigated development practices by not 
constraining to specific organization (e.g., Microsoft as in \cite{Meyer-CharacterizeSoftDevPerceptionOfProductivity-ESEM2017}). 
Each participant was asked to provide feedback on the survey questions. 
Each interview session lasted about half an hour. At the end of the interview sessions, we analyzed how the interview participants 
mentioned about different stakeholders while discussing their expectations. We formulated one survey question per identified 
stakeholder. For example, one stakeholder is the software organization itself, i.e., where the developer worked. As such, our first 
survey question as: \emt{What is expectation from your organization?}. We then checked whether we could use existing literature 
on developer productivity and well-being to provide a clear answer to the survey question. For the above survey question, while we can take 
cues from existing literature, we found that the literature focused on specific use case scenarios within an organization (e.g., interruption management) and 
not on the diverse types of expectations that developers may have towards their organization. We thus kept the survey question. At the end, 
we formulated six survey questions, each pointing toward a specific stakeholder, e.g., organization, manager, peer, new hires, university and government.
In \tbl\ref{table:questions_mapping}, we show the survey question. 
\begin{table}[h]
    \centering
    
    \caption{Questions asked in our survey and their mapping to our Research Questions (RQ) in \sec\ref{sec:results}}
	\resizebox{\columnwidth}{!}{%
	    \begin{tabular}{lll}
    \toprule
        \textbf{No.} & \textbf{Question} & \textbf{RQ} \\ \midrule
        1 & What are your expectations from your organization? & 1 \\
        2 & What are your expectations from your manager? & 2 \\
        3 & What are your expectations from peers in the team? & 3 \\
        %4 & What are your expectations from your peers in the team? & RQ3 \\
        4 & What are your expectations from the new hires? & 4 \\
        5 & What are your expectations from the universities? & 5 \\
        6 & What are your expectations from the government? & 5 \\
        \bottomrule
    \end{tabular}%
    }
    \label{table:questions_mapping}
    \vspace{-3mm}
\end{table}
\subsection{Conduct Survey}
\label{survey_participants}
% \input{Tables/Survey_Questions_1}
% The survey questions are shown in Table \ref{table:survey_questions_1}.
For our survey, we targeted professional software developers working in small, medium, and large 
organizations. To select the survey participants, we combined judgmental sampling~\cite{Vogt2005} with snowball sampling. In judgemental sampling, we start with a selected list of participants who we know that could answer our survey questions properly and who fit well within our target population (e.g., software developers working in teams). In snowball sampling~\cite{creswell2013}, we ask the participants who already answered to the survey question to recommend our survey to other suitable and potential participants. Given the nature of snowball sampling, it is not possible to keep track of how many actual invitations are sent. As such, we cannot report the response ratio. However, snowball sampling ensures that we reach to a large variety of suitable participants, whom we could not possibly have targeted via other means of communication. For our case, the combination of judgemental and snowball sampling work better than other means like posting the survey invitations online and expecting anyone/everyone to respond to the surveys. As we can imagine, such online posting could attract responses from unsuitable developers and it would have been difficult for us to ensure quality in the survey responses.

\begin{table}[h]
\centering
\caption{Experience wise Distribution of Participants}
\begin{tabular}{lr}
\toprule
\textbf{Role of the Participants} & \textbf{Percentage}\\
\midrule
less than 2 years & 34.8\%\\ 
2 to 5 years & 28.2\%\\ 
5 to 10 years & 18.2\%\\
more than 10 years & 15.5\%\\
experience not disclosed & 3.3\%\\
\bottomrule
\end{tabular}
\label{tab:experience}
 \vspace{-3mm}
\end{table}
\begin{table}[h]
\centering
\caption{Organization size wise Distribution of Participants}
\begin{tabular}{llr}
\toprule
\textbf{Size Category} & \textbf{Organization Size}  & \textbf{No. of Respondents}\\
\midrule
Small&  1 - 50  & 39 \\
Small to medium & 51 - 150 & 73 \\
Medium & 151 - 500 & 15  \\
Large & 500+ & 9  \\
\bottomrule
\end{tabular}
\label{tab:org_size}
\end{table}

We conducted the survey in two phases, i.e., we repeated our judgemental and snowball sampling approach twice. This is to ensure that we could cover as many potential participants as possible. The two phases were open for two months in the last quarter of 2020. We received total 181 responses (136 from first survey and 45 from the second survey). Each participant in the survey was first asked a series of demographic questions (e.g., roles, experience, gender) and
then was presented the survey questions related to the expectation in software development practices. In total, we asked six questions to a participant (see \tbl\ref{table:questions_mapping}). All the questions are open-ended, i.e., the participants were asked to write as much as possible in texts to answer to each survey question.

Among the survey respondents, 17.6\% developers are team leads and the rest noted various development roles (e.g., developer, data engineer, software quality assurance engineer, software architects, etc.). In \tbl\ref{tab:experience}, we show the distribution of the survey respondents by their years of experience. Around 50\% of the survey respondents have up to 5 years of software development experience in the industry, which around 46\% of the respondents have more than 5 years of professional software development experience. Therefore, our survey responses contain a balanced mix of experienced, experts, and novice professional software developers. In \tbl\ref{tab:org_size}, we show the distribution of the survey respondents based on the size of the organizations, i.e, the companies they worked on during the time of our survey. We find a good concentration of respondents from small and medium sized organizations, while few responses from the large organizations. As such, our analysis of the survey can be more applicable to small or medium sized companies. This is not a bad thing, given we already have substantial information on the productivity measures of developers in big companies like IBM~\cite{Albrecht-MeasureApplicatonDevelopmentProductivity-IBM1979,Paiva-FactorsProductivitySoftDev-CSSE2010} or Microsoft~\cite{Meyer-CharacterizeSoftDevPerceptionOfProductivity-ESEM2017}, but not so much on the small or medium-sized companies. Our study may guide more on the management of expectations in small or medium sized companies.

\subsection{Survey Data Analysis}
\label{survey_data_collection}
All of our survey questions are open-ended, for which we applied an open coding approach~\cite{miles_1994} which includes labelling of concepts/categories in textual contents based on the properties and dimensions of the entities about which the contents are provided. In our open coding, we followed the card sorting approach~\cite{Fincher2005}.
In card sorting, the textual contents are divided into cards, where each card denotes a conceptually coherent quote. 
For example, consider the following sentence: “Well Behaved and Proper Project Timeline.” The sentence has two different conceptual coherent quotes, “Well Behaved”, and “Proper Project Timeline”. The first quote refers to good behavior. The second quote refers to the proper management. In our analysis, as we analyzed the quotes, themes and categories emerged and evolved during the open coding process.

\begin{table}[t]
\caption{The agreement level in the open
coding sessions between the two coders (\#Q denotes \# of quotes used)}
\label{table: agreement_level}
\centering
\begin{tabular}{lp{1cm}p{1cm}p{1cm}p{1cm}p{1cm}}
\toprule
            %   \multirow{2}{*}{} & \textbf{RQ1} & \textbf{RQ2} & \textbf{RQ3} & \textbf{RQ4} & \textbf{RQ5} \\ %\hline
            %     & (\#Q = 79) & (\#Q = 59) & (\#Q = 94) & (\#Q = 66) & (\#Q = 107) \\
                & \textbf{RQ1 (\#Q79)} & \textbf{RQ2 (\#Q59)} & \textbf{RQ3 (\#Q94)} & \textbf{RQ4 (\#Q66)} & \textbf{RQ5 (\#Q107)} \\ %\hline
                %& (\#Q = 79) & (\#Q = 59) & (\#Q = 94) & (\#Q = 66) & (\#Q = 107) \\
                \midrule
%{Quotes} & 74          & 73          & 74          \\ %\hline
Percent         & 76.2         & 78.3         & 73.7        & 77.4        & 70.5 \\ %\hline
Cohen   $\kappa$        & 0.73        & 0.754        & 0.71       & 0.749         & 0.67  \\ %\hline
Scott  $\pi$         & 0.73        & 0.754        & 0.71       & 0.749         & 0.67  \\ %\hline
Krippen  $\alpha$       & 0.731        & 0.755        & 0.711      & 0.75         & 0.671  \\ %\hline

\bottomrule
\end{tabular}
 \vspace{-4mm}
\end{table}We analyzed the survey responses in four phases as outlined below.
First, the first two authors independently coded the responses of each question to extract potential categories. Second, the
authors conducted discussion sessions to develop a unified common coding scheme
for each question using these categories. Third, the responses were
coded using the coding scheme developed from the survey. This approach resulted in a set of quotes per survey response and one or more assigned label as expectation type for each quote. Fourth, the last author was consulted to refine and and finalize the expectation types and to cluster the types into higher categories.  The level of agreement between the first two coders is presented in Table \ref{table: agreement_level} using four measures (using online tool Recal2~\cite{Recal2}): 1) Percent agreement, 2) Cohen $\kappa$~\cite{Cohen1960}, 3) Scott’s $\pi$~\cite{scott1955},
and 4) Krippendorff’s $\alpha$~\cite{krippendorff2004}.  The average
$\kappa$ value was 0.723 and Krippen $\alpha$ value if 0.724. To consider a $\kappa$ value between 0.61 and 0.80 \cite{Landis1977} as a `substantial agreement’ is a common practice.

\section{Study Results}\label{sec:results}
In this section, we answer five research questions:
\begin{enumerate}[label=\bf{RQ\arabic{*}.}, leftmargin=30pt]
  \item What do professional software developers in our survey expect from their organizations? (\sec\ref{sec:rq1})
  \item What do professional software developers in our survey expect from managers? (\sec\ref{sec:rq2})
  \item What professional software developers in our survey expect from peers in the team? (\sec\ref{sec:rq3})
  \item What professional software developers in our survey expect from the new hires? (\sec\ref{sec:rq4})
  \item What do the developers in our survey expect from government and universities? (\sec\ref{sec:rq5})
\end{enumerate} \begin{table*}[h]
    \centering
    \caption{The expectation types observed during the open-coding of the responses (\#TC = Total Categories, \#TR = Total number of respondents, \#C = the number of code, \#R = the number of respondents)}
    %\resizebox{\columnwidth}{!}{%
    \begin{tabular}{l|l l|l l|l l|l l|l l|l l}
        \toprule
       \bf{Expectation} & \multicolumn{2}{c}{RQ1} & \multicolumn{2}{c}{RQ2} & \multicolumn{2}{c}{RQ3} & \multicolumn{2}{c}{RQ4} & \multicolumn{2}{c}{RQ5} & \multicolumn{2}{c}{} \\ 
       \cmidrule{2-11}
        \bf{Type} & \#C & \#R & \#C & \#R & \#C & \#R & \#C & \#R & \#C & \#R & \#TC & \#TR \\ \midrule
        
\bf{Work Benefits} & 14 & 12 &   &  &   &  &   &  &   &  & 14 & 12 \\ \cmidrule{2-13}
\bf{Goal Achievement} & 4 & 4 &   &  &   &  &   &  &   &  & 4 & 4 \\ \cmidrule{2-13}
\bf{Career Opportunities} & 11 & 9 & 26 & 22 &   &  &   &  & 17 & 16 & 54 & 47 \\ \cmidrule{2-13}
\bf{Work-place Culture} & 24 & 20 &   &  &   &  &   &  &   &  & 24 & 20 \\ \cmidrule{2-13}
\bf{Learning Opportunities} & 3 & 3 &   &  &   &  &   &  &   &  & 3 & 3 \\ \cmidrule{2-13}
\bf{Proper Management} & 15 & 13 & 58 & 46 &   &  &   &  &   &  & 73 & 59 \\ \cmidrule{2-13}
\bf{Knack in Standard SE Practice} & 5 & 5 &   &  & 64 & 46 & 19 & 16 &   &  & 88 & 67 \\ \cmidrule{2-13}
\bf{People Skill} &   &  & 65 & 51 &   &  &   &  &   &  & 65 & 51 \\ \cmidrule{2-13}
\bf{Tech Savvy} &   &  & 14 & 14 &   &  &   &  &   &  & 14 & 14 \\ \cmidrule{2-13}
\bf{Industry Oriented Teaching} &   &  &   &  &   &  &   &  & 72 & 54 & 72 & 54 \\ \cmidrule{2-13}
\bf{Learning Environment} &   &  &   &  & 9 & 9 &   &  & 22 & 15 & 31 & 24 \\ \cmidrule{2-13}
\bf{Career Support} &   &  &   &  &   &  &   &  & 25 & 19 & 25 & 19 \\ \cmidrule{2-13}
\bf{Sincerity} &   &  &   &  & 57 & 37 & 24 & 16 &   &  & 81 & 53 \\ \cmidrule{2-13}
\bf{Motivation for Learning} &   &  &   &  &   &  & 22 & 18 &   &  & 22 & 18 \\ \cmidrule{2-13}
\bf{Policy and Facility} &   &  &   &  &   &  &   &  & 56 & 41 & 56 & 41 \\ \cmidrule{2-13}
\bf{Proper Education} &   &  &   &  &   &  &   &  & 18 & 15 & 18 & 15 \\ \cmidrule{2-13}
\bf{Support for Industry} &   &  &   &  &   &  &   &  & 20 & 19 & 20 & 19 \\ \cmidrule{2-13}
\bf{Supportive Attitude} &   &  &   &  & 61 & 47 &   &  &   &  & 61 & 47 \\ \midrule
\bf{Total} & 76 & 66 & 163 & 133 & 191 & 139 & 65 & 50 & 230 & 179 & 725 & 567 \\ \midrule
\bf{\# of Expectation Types per RQ} & \multicolumn{2}{c}{7} & \multicolumn{2}{c}{4} & \multicolumn{2}{c}{4} & \multicolumn{2}{c}{3} & \multicolumn{2}{c}{7} & \multicolumn{2}{c}{18} \\
        
        \bottomrule
    \end{tabular}%
   % }
    \label{table:consolidated_categories}
\end{table*} 
The last column in \tbl\ref{table:questions_mapping} shows how our survey questions are mapped to the five research questions (RQ). Our open coding of the survey responses resulted in a total of 18 expectation types across the five research questions. In \tbl\ref{table:consolidated_categories}, we show the number of quotes and survey respondents for each expectation type as we found in our survey data. In the following, we explain the expectation types by each RQ. Each expectation type is shown as $Type^{\#NQ,\#NR}$, where \#NQ is the number of quotes generated for the expectation type and \#NR is the number of type related responses. We use $R_{i,j}$ for the respondent identity in the example of quotes where $i$ denotes survey phase id (1 or 2) and $j$ denotes the numeric ID of the respondent. The graph bars are annotated with the number of category specific responses.
% \gias{the result section for each RQ is still quite shallow. Please enhance as follows. Discuss each category (e.g., one para per seven categories in Fig. 1) as a separate para. Follow this format: para title (with two superscripts, number of quotes, number of users). then give an overview what this category denotes for the RQ. summarize the key finidngs by first explaining the simple text, then show two/three quotes from the respondder. For each repsponder, identify the respondeder by this: $R_{i,j}$ where $i$ denotes survey id (e.g, 1 or 2), $j$ denotes the numeric ID of the responder (e.g., 1, 2, 3,...). For each category, if you can compare the finding with previous study findings, that would be great.}

\subsection{Expectations from Organizations (RQ1)}\label{sec:rq1}

\subsubsection{Motivation}
Employees always have expectations from their organization, and being satisfied with those make them loyal to the company. The expectations vary from profession to profession. When it comes to expectations in the workplace, the employers and the workers may have different points of view. Sometimes, the differences are so huge that it affects the whole organization. For software companies, understanding the expectations of employees is important as software developers usually have easy job switch opportunities, and companies can suffer from turnovers. 
% If an organization wants to improve employee engagement and retention, he/she will need both a deeper understanding of employee expectations and modern business best practices that fulfill them.
%But, having an imbalance between the organization expectations and employee needs is quite common in new age companies today. Business professionals and entrepreneurs these days are facing a challenge regarding getting their job done while fulfilling their employees' needs. 

%Bringing a balance between the needs and expectations of both the employee and employer can solve the problem most of the time. But before that, we need to know what causes the problem. 

\begin{figure}[h]
\centering
\includegraphics[scale=0.14]{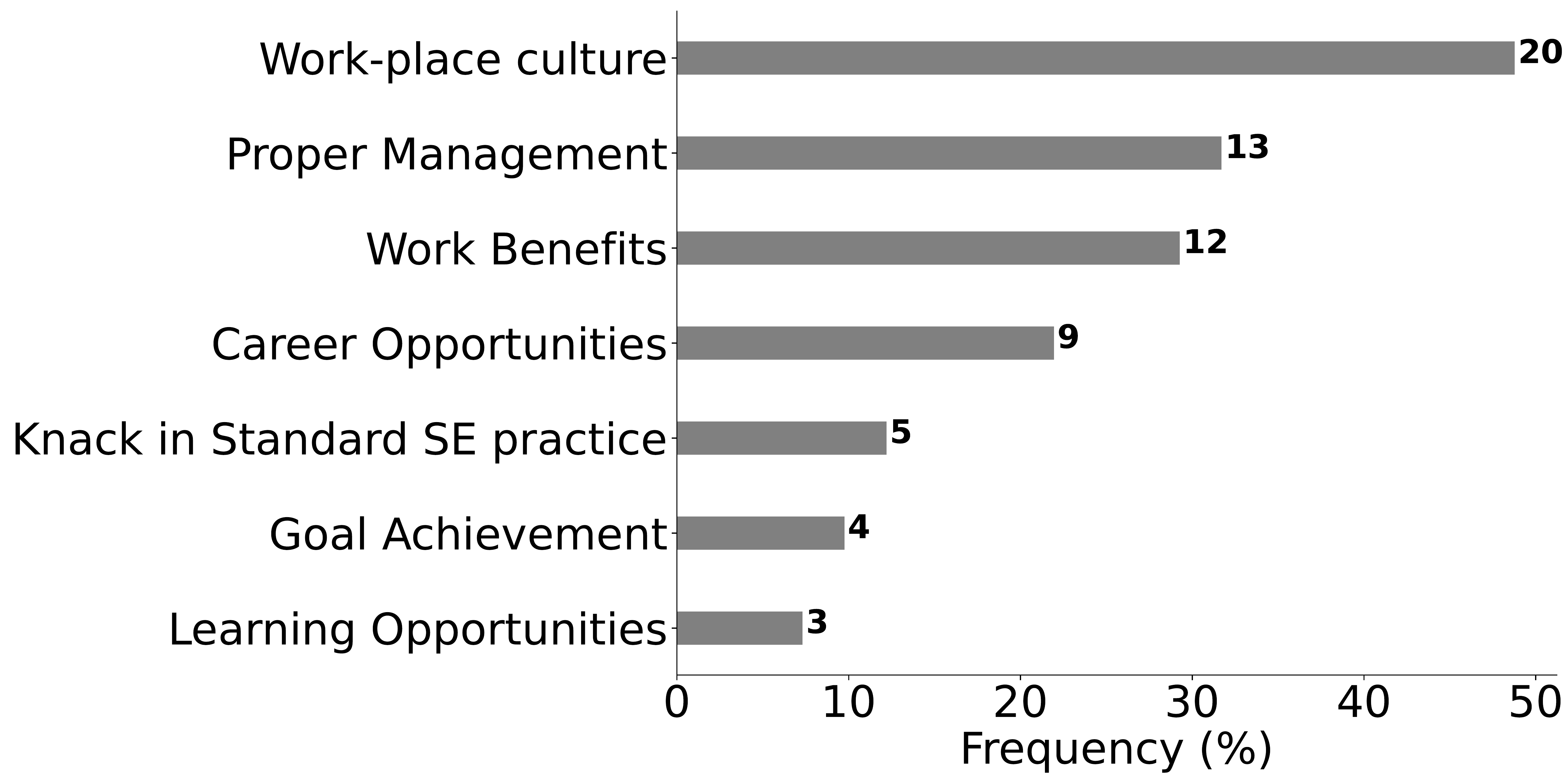} 
\caption{Expectations from organization}
\label{fig:expectations_from_organization}
\vspace{-5mm}
\end{figure}

\subsubsection{Results}
% \gias{can you make the figures black and white? also can you put actual numbers at the end of each bar?} \khalid{updated. is it okay to draw bar frequency-wise and annotate response number?}\gias{Yes, it's fine, put all bars similar width across all figures. also make the response number fonts much bigger - also remove the right and top border from each bar chart.}
\fig\ref{fig:expectations_from_organization} summarizes seven expectation types as found in our survey.
\begin{inparaenum}[(1)]\item $Workplace$ $Culture^{24,20}$. Participants expect their organizations to focus on workplace culture (48.78\%) that covers overtime-free and politics-free environment, work-life balance, good relations, etc. $R_{2,24}$ mentions \textit{``A friendly work environment with the best practice of modern methodologies''}. 
\item $Proper$ $Management^{15,13}$. Participants (31.71\%) expect proper guidance, judgment, recognition, etc. $R_{2,12}$ mentions \textit{``To give me proper guidance in order to perform my duty''}. 
\item $Work$ $Benefits^{14,12}$. Participants (29.27\%) expect timely salary, proper salary, and food supply. $R_{2,17}$ expects \textit{``giving salary according to contribution''}. 
\item $Career$ $Opportunities^{11,9}$. 21.95\% of the participants expect various career opportunities including career growth and security, proper structure from the organization. The respondents expect a proper career path within the organization, for example, the respondent $R_{2,2}$ mentions this, \textit{``Introduce a proper structure which will benefit both employees and the organization''}.
Other expectations types are  \item $Adoption$ $of$ $Standard$ $SE$ $practice^{5,5}$ (12.2\%), \item proper support for $Goal$ $Achievement^{4,4}$ (9.76\%), and \item $Learning$ $Opportunities^{3,3}$ (7.32\%).
\end{inparaenum}

\begin{tcolorbox}[flushleft upper,boxrule=1pt,arc=0pt,left=0pt,right=0pt,top=0pt,bottom=0pt,colback=white,after=\ignorespacesafterend\par\noindent]
 \bf{RQ$_1$ Expectations from Organizations.} 
 The major expectation from the organization is a thriving and supportive workplace culture. Proper management, work benefits, and career opportunities for the employees are also expected.
\end{tcolorbox}
\subsection{Expectations from Manager (RQ2)}\label{sec:rq2}
\subsubsection{Motivation} 
The managers are mainly responsible for coordinating the team members and the client, and streamlining the development process by properly allocating the resources. They can also actively participate in the development. To efficiently manage the software development process, the manager must understand the expectations of all the team members towards him/her. 
% This understanding will play a crucial role in the successful completion of the development process.

\begin{figure}[h]
\centering
\includegraphics[scale=0.14]{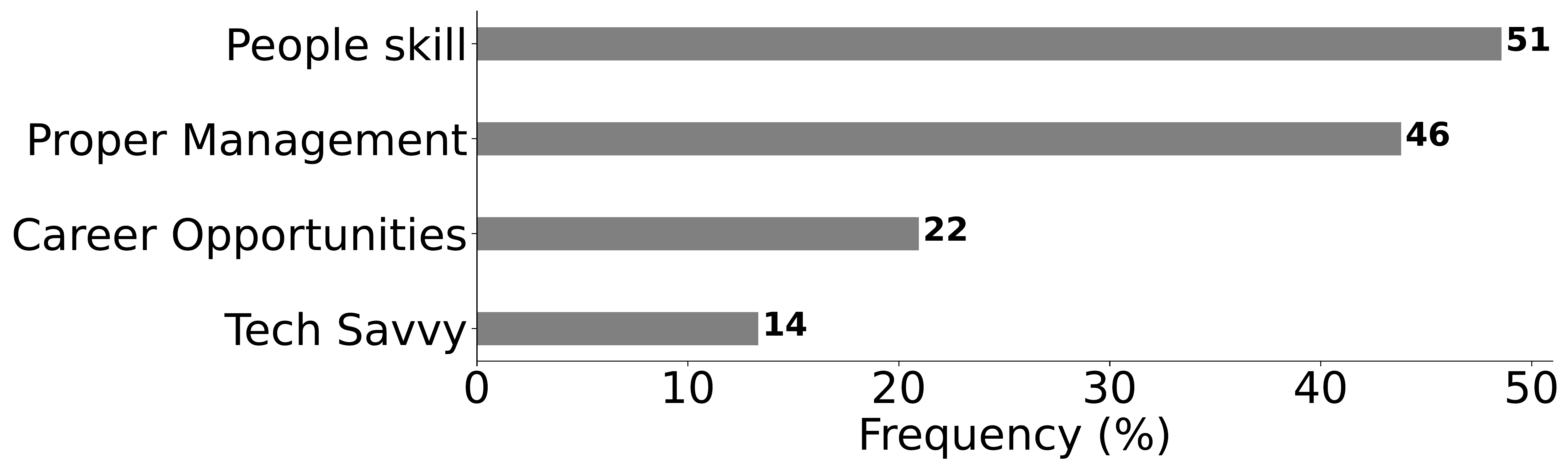} 
\caption{Expectations from Manager}
\label{fig:expectations_from_manager}
\vspace{-5mm}
\end{figure}
\subsubsection{Results}
\fig\ref{fig:expectations_from_manager} shows the dominant expectations from manager.
\begin{inparaenum}[(1)]
\item $People$ $skill^{65,51}$. Participants (48.57\%) want their manager to have people skill implying that managers need to understand and guide their teammates. $R_{2,24}$ expected \textit{``To guide us to solve our problems and also think about what might impact our career''}. $R_{2,29}$ expected \textit{``Proper guidance for career growth''}. 
\item $People$ $Management^{58,46}$. 43.81\% of the participants expect managers to have proper accountability, clear vision, reasonable deadline, etc. $R_{2,42}$ expects \textit{``Future proof vision, good leading qualities''}. 
\item $Career$ $Opportunities^{26,22}$. 20.95\% of participants want their manager to be helpful in their career opportunities by providing recognition, training, etc. They want proper recognition from their managers, namely from the respondent $R_{2,22}$ we can find this, \textit{``Proper communication with the employees, Recognition of work''}.
\item $Tech$ $Savvy^{14,14}$. They are also expected to be technologically proficient and use advanced tools as per 13.33\% of the respondents. The respondents want their manager to be comfortable with the technical difficulties and latest technologies and tools, for instance, one participant $R_{2,27}$ noted \textit{``Understand technical difficulties when inform them about it''}, and another one $R_{2,43}$ noted \textit{``Would approach more latest tools and techs to keep updated with trends''}.
\end{inparaenum}

\begin{tcolorbox}[flushleft upper,boxrule=1pt,arc=0pt,left=0pt,right=0pt,top=0pt,bottom=0pt,colback=white,after=\ignorespacesafterend\par\noindent]
 \bf{RQ$_2$ Expectations from Manager.}
The major expectation from the managers is their people skill and management. The managers are also expected to be tech savvy and help the employees with their career opportunities.
\end{tcolorbox}
\subsection{Expectation from Peers (RQ3)}\label{sec:rq3}
\subsubsection{Motivation} 
Software development is very much a teamwork. The developers use APIs developed by peers, review codes written by others. They also have to collaborate with the quality assurance team, DevOps team, implementation and support team. Having good working relations and understanding with peers is likely to improve a team's productivity to a great extent. On the other hand, issues between peers have a significant impact on the peers. Hence, understanding the expectation from peers is important to the employees as well as managers. 
% All can play their roles to develop a good team spirit if they are informed of the expectations.

\begin{figure}[h]
\centering
\includegraphics[scale=0.15]{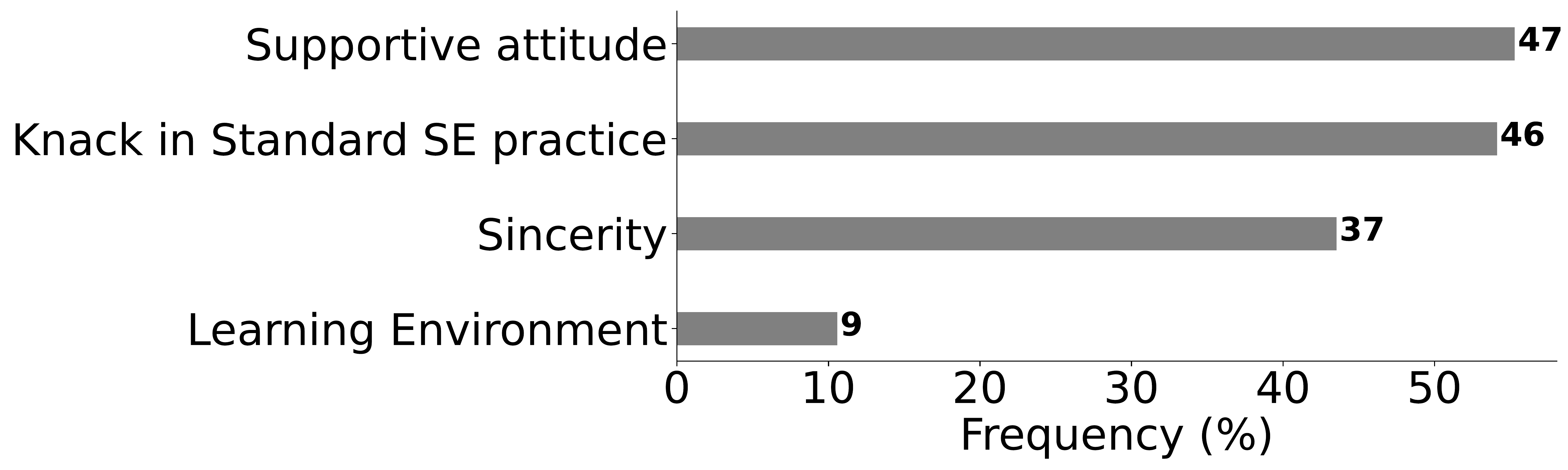} 
\caption{Expectations from Peers}
\label{fig:expectations_from_peers}
\vspace{-5mm}
\end{figure}
\subsubsection{Results}
%To see the expectations of employees from their peers, we have reported some pivotal points at the 
Figure~\ref{fig:expectations_from_peers} shows the distribution of four expectation types from peers.
\begin{inparaenum}
\item $Supportive$ $Attitude^{61,47}$. Participants dominantly expect their peers to have supportive attitude (55.29\%) which comprises trustworthiness, honesty, good relationship, clear communication, good behavior, etc. $R_{2,7}$ expects peers to \textit{``being helpful about technical knowledge''}.
\item $Knack$ $in$ $Standard$ $SE$ $practice^{64,46}$. Moreover, they need to have knack in software engineering (SE) practices as per the expectation of 54.12\% participants including quality coding practice, time management, etc. They also want their peers to write reusable codes (e.g., $R_{2,11}$).
\item $Sincerity^{57,37}$. The peers are also expected  (by 43.53\% participants) to be sincere in passion and responsibility at work. $R_{2,12}$ expects from peers \textit{``That they obey by the company rules''}. The respondents also want them to work hard enough to manage everything. $R_{2,4}$ expected \textit{``Working hard to manage everything quite easily''}.
\item $Learning$ $Environment^{9,9}$. On top of that, they are expected to be involved in a learning environment (10.59\%) by being a learner, sharing knowledge, etc. The respondents want them to pursue their knowledge and interested to be familiar with new technology. For instance, $R_{2,41}$ noted this, \textit{``eagerness to learn new technologies''}. $R_{2,38}$ noted this, \textit{``Should always learn to get familiar with new technology''}.
\end{inparaenum}
\begin{tcolorbox}[flushleft upper,boxrule=1pt,arc=0pt,left=0pt,right=0pt,top=0pt,bottom=0pt,colback=white,after=\ignorespacesafterend\par\noindent]
 \bf{RQ$_3$ Expectations from Peers.}
The major expectation from the peers is their supportive attitude and knack in standard SE practices to write high-quality reusable code. The peers are expected to be sincere and keen to learn new technologies.
\end{tcolorbox}
\subsection{Expectation from New Hires (RQ4)}\label{sec:rq4}

\subsubsection{Motivation}
The new hires in a company are not likely to have clear idea about the norms and conventions, standards and practices of a software company. It is the responsibility of the company to make these clear. However, many start-ups or small organizations do not have strong non-tech departments and often fails to communicate the expectations from their new hires. Consequently, new recruits build up a different point of expectations themselves which may eventually affect the growth of the organization. %To bridge this gap, we need to have a proper understanding of new hires' expectations. 

% Hence, this circumstance has motivated us to find the expectations from the new recruits for the well-being of the software industry.
\begin{figure}[h]
\centering
\includegraphics[scale=0.15]{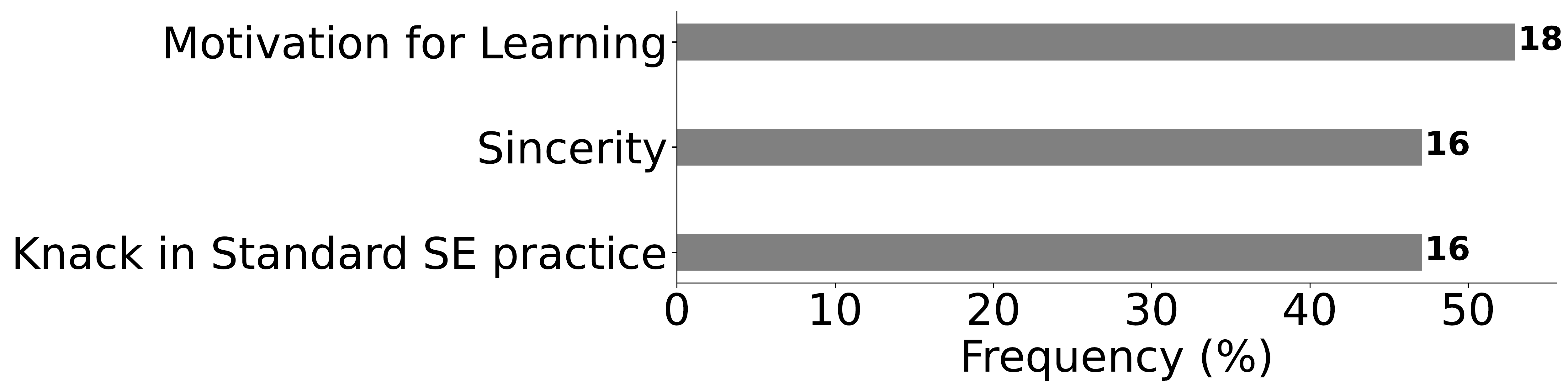} 
\caption{Expectations from New Hires}
\label{fig:expectations_from_new_hires}
\vspace{-5mm}
\end{figure}

\subsubsection{Results}
\fig\ref{fig:expectations_from_new_hires} summarizes the three expectation types we observed from the new hires. 
\begin{inparaenum}[(1)]
\item $Motivation$ $for$ $Learning^{22,18}$. We have found that the recruits are expected to focus mostly on their learning as per most of the respondents (52.94\%). In detail, they have to be enthusiastic, quick learner with capability of better thinking, self-improvement, etc. The participants consider the new hires as self-learner and be adaptive, for instance, the respondent $R_{2,15}$ wrote this, \textit{``Nobody expects you to know all the technology, just learn to adapt and learn new technology''}. The recruits are also supposed to be interested in self-improvement, namely the respondent $R_{2,44}$ wrote down this, \textit{``Has to be open to self improvement for personal and professional gain''}
\item $Knack$ $in$ $Standard$ $SE$ $Practice^{19,16}$. Moreover, they need to be SE practitioners (47.06\%) which covers professional coding practice, problem-solving attitude, feeling of team player, etc. The respondents think that the recruits should have basic knowledge about the technology and CS fundamentals and be able to write clean code, for example, the respondent $R_{2,39}$ mentioned \textit{``Strong basic knowledge about the technology''}, the respondent $R_{2,23}$ mentioned \textit{``Having basic knowledge of CS fundamentals''}, and the respondent $R_{2,2}$ noted \textit{``Learn how to code cleanly''}.
\item $Sincerity^{24,16}$. The new hires are expected to maintain their sincerity (47.06\%) which encompasses transparency, co-operation, solidarity, etc. The respondents want them to have good behavior, particularly the respondent $R_{2,44}$ wrote down this, \textit{``Has to be a good person and others have to be able to feel that''}. The new hires are also expected to follow the company rules sincerely, particularly the respondent $R_{2,12}$ \textit{``That they obey by the company rules sincerity''}.
\end{inparaenum}

\begin{tcolorbox}[flushleft upper,boxrule=1pt,arc=0pt,left=0pt,right=0pt,top=0pt,bottom=0pt,colback=white,after=\ignorespacesafterend\par\noindent]
 \bf{RQ$_4$ Expectations from New Hires.}
The major expectation from the new hires is their motivation for quick learning with better thinking and self improvement. The new hires are expected to be sincere and have a knack in standard SE practices.
\end{tcolorbox}
\subsection{Expectation from Government \& Educational Institutions (RQ5)}\label{sec:rq5}
\subsubsection{Motivation} 
Government plays an important role to support the tech industry. However, the steps taken by the government can lag behind the expectation of practitioners. Moreover, to have a real impact on the growth of the SE industry, it is required to know exactly where  government intervention is required and how the valuable public money should be best utilized. Universities play an important role in the SE industry by providing qualified engineers. However, to play their role properly, they need to know whether their graduates can fulfill the current demands in the SE industry. Also, what the graduates think in retrospect about the training they received and if they feel any scope of improvement are also important for the continuous improvement of the university. 
% Thus we need to know what are the expectation of software practitioners from the government and the universities.
\begin{figure}[h]
\centering
\includegraphics[scale=0.15]{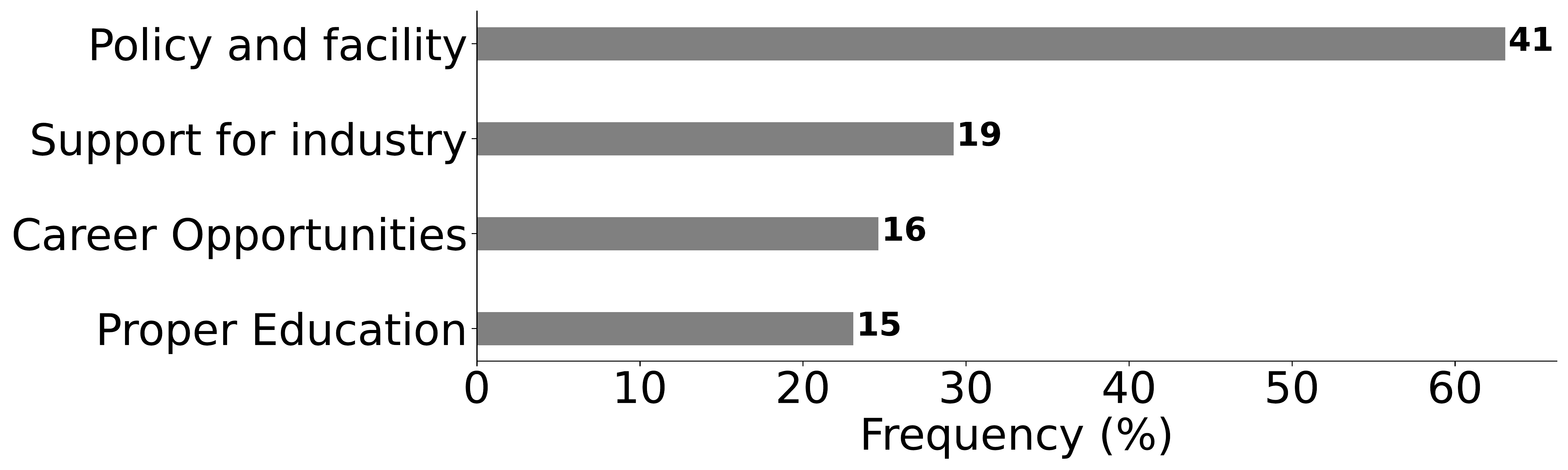} 
\caption{Expectations from government}
\label{fig:expectations_from_government}
\vspace{-5mm}
\end{figure}
\begin{figure}[h]
\centering
\includegraphics[scale=0.15]{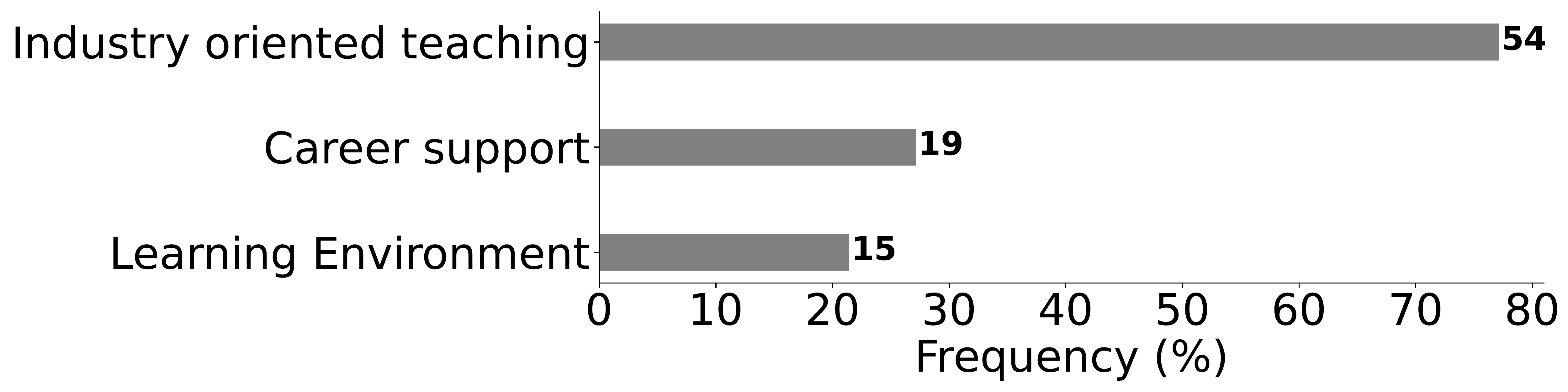} 
\caption{Expectations from universities}
\label{fig:expectations_from_universities}
\vspace{-3mm}
\end{figure}

\subsubsection{Results} 
The expectations of software practitioners from governments and universities are presented in Figures~\ref{fig:expectations_from_government}, and~\ref{fig:expectations_from_universities} respectively.
\begin{inparaenum} 
\item $Policy$ $and$ $facility^{56, 41}$. Practitioners in most cases (63.08\%) expect various job-related policies and facilities from the government. They expect a separate entity for computer science and engineering (CSE) professional like other engineering disciplines. $R_{2,34}$ noted this, \textit{``Positive move, Introduced CSE as an engineering wing as like others (civil, architect, etc.) in the state''}. The facilities include high-speed internet, a simple taxation process, etc. 
They expect a separate zone for software firms including all the facilities, particularly the respondent $R_{1,108}$ mentioned this, \textit{``Well facilitated IT parks near/in each major cities.''}. 
\item $Support$ $for$ $industry^{20, 19}$. Respondents expect special support from the government  to let the industry grow. The supports includes implementation of regulations regarding SE industry. $R_{2,26}$ mentioned \textit{``Government should acknowledge this growing market properly and consider some projects and rules regulations to help growing this market fast and rank high in world.''}. 
% Moreover, the respondents demands priority of local vendors in government purchase to support the industry as $R_{2,5}$ noted \textit{``Don't always buy software from foreign vendors. Create a separate identity for software engineers.''}.
\item $Career$ $Opportunities^{17,16}$. Respondents expect the government to be an active participants in creating job opportunities. $R_{2,11}$ mentioned \textit{``1. Will create more working opportunities for software engineer 2. Will create more government employment 3. Will take steps for ensuring high speed internet in low costs.''}. 
\item $Proper$ $Education^{18,15}$. Respondents expect government supporting in implementing a standard up-to-date curriculum in CS education. $R_{2,31}$ wrote \textit{``Supports to company and universities to create improved curriculum that help to add value to Software industries.''}
\item $Industry$ $Oriented$ $Teaching^{72,54}$. Practitioners mostly (77.14\%) expect \textit{industry oriented teaching} from the universities. $R_{2,39}$ noted this, \textit{``Few industry related course, make student familiar with new technologies which are being used in software industries''}.
\item $Career$ $Support^{25,19}$. Respondents have mentioned that they expect careers related support such as job fairs, internships, industry visits from the universities. $R_{2,17}$ wrote this \textit{``Arrange seminar, talk through which current students can get idea of software industry life from alumni.''}.
\item $Learning$ $Environment^{22,15}$. Respondents expect an overall learning environment from the universities. According to respondents, universities should focus on skill development, teamwork, and communication skills. 
% Respondent $R_{2,26}$ mentioned \textit{``Universities should emphasize team work and team communication, industrial cultural knowledge along with quick learning ability and egger to update themselves with new technology.''}
\end{inparaenum}

\begin{tcolorbox}[flushleft upper,boxrule=1pt,arc=0pt,left=0pt,right=0pt,top=0pt,bottom=0pt,colback=white,after=\ignorespacesafterend\par\noindent]
 \bf{RQ$_5$ Expectations from Government and Universities.}
The major expectation from the government is SE industry-friendly policies and facilities, and from the university is industry-oriented teaching.
\end{tcolorbox}

% \newcolumntype{b}{X}
\newcolumntype{v}{>{\hsize=.04\hsize}X}
\newcolumntype{k}{>{\hsize=.96\hsize}X}
\newcolumntype{m}{>{\hsize=.2\hsize}X}
\newcolumntype{y}{>{\hsize=.33\hsize}X}
\begin{table*}[!ht]
    \centering
    \caption{Highlights of Findings from Survey Questions by Experience}
    \begin{tabularx}{\textwidth}{v|k}
        \hline
        RQ1 & What are your expectations from your organization?
        \newline Career Opportunities = CO, Goal Achievement = GA, Knack in Standard SE Practice = SP, Learning Opportunities = LO, \newline Proper Management = PM, Work Benefits = WB, Work-place Culture = WC\newline
        1) \textbf{less than 2}: CO (12.5\%), GA (12.5\%), SP (12.5\%), LO (6.25\%), PM (50.0\%), WB (31.25\%), WC (43.75\%), 2) \textbf{2 to 5}: CO (31.25\%), GA (6.25\%), SP (12.5\%), LO (6.25\%), PM (18.75\%), WB (37.5\%), WC (56.25\%), 3) \textbf{greater than 5}: CO (22.22\%), GA (11.11\%), SP (11.11\%), LO (11.11\%), PM (22.22\%), WB (11.11\%), WC (44.44\%)
        \\ \hline
        
        RQ2 & What are your expectations from your manager?\newline Career Opportunities = CO, People Skill = PS, Proper Management = PM, Tech Savvy = TS\newline 
        1) \textbf{less than 2}: CO (21.62\%), PS (54.05\%), PM (32.43\%), TS (10.81\%), 2) \textbf{2 to 5}: CO (11.76\%), PS (52.94\%), PM (50.0\%), TS (14.71\%), 3) \textbf{greater than 5}: CO (29.41\%), PS (38.24\%), PM (50.0\%), TS (14.71\%)
        \\ \hline
        
        RQ3 & What are your expectations from peers in the team? \newline Knack in Standard SE Practice = SP, Learning Environment = LE, Sincerity = SC, Supportive Attitude = SA\newline 
        1) \textbf{less than 2}: SP (29.63\%), LE (3.7\%), Sincerity (22.22\%), SA (85.19\%), 2) \textbf{2 to 5}: SP (59.09\%), LE (22.73\%), Sincerity (36.36\%), SA (45.45\%), 3) \textbf{greater than 5}: SP (69.44\%), LE (8.33\%), Sincerity (63.89\%), SA (38.89\%)
        \\ \hline
        
        RQ4 & What are your expectations from the new hires?\newline
        Knack in Standard SE Practice = SP, Motivation for Learning = ML, Sincerity = SC\newline 
        1) \textbf{less than 2}: SP (50.0\%), ML (66.67\%), Sincerity (66.67\%), 2) \textbf{2 to 5}: SP (57.14\%), ML (35.71\%), Sincerity (21.43\%), 3) \textbf{greater than 5}: SP (25.0\%), ML (62.5\%), Sincerity (62.5\%)
        \\ \hline
        
        RQ5 & What are your expectations from the universities and government?\newline
        Universities: Career Support = CS, Industry Oriented Teaching = IOT, Learning Environment = LE\newline
        1) \textbf{less than 2}: CS (26.32\%), IOT (57.89\%), LE (52.63\%), 2) \textbf{2 to 5}: CS (27.78\%), IOT (88.89\%), LE (11.11\%), 3) \textbf{greater than 5}: CS (27.27\%), IOT (81.82\%), LE (9.09\%)
        %\\ %\hline
        \newline
        %RQ5 & What are your expectations from the Government?\newline
        Government: Career Opportunities = CO, Policy and Facility = PF, Proper Education = PE, Support for Industry = SI\newline
        1) \textbf{less than 2}: CO (33.33\%), PF (55.56\%), PE (22.22\%), SI (27.78\%), 2) \textbf{2 to 5}: CO (26.67\%), PF (66.67\%), PE (13.33\%), SI (33.33\%), 3) \textbf{greater than 5}: CO (18.75\%), PF (65.62\%), PE (28.12\%), SI (28.12\%)
        \\ \hline
    \end{tabularx}
    \label{table:analysis_by_experience_part1}
\end{table*}
% \newcolumntype{b}{X}
\newcolumntype{v}{>{\hsize=.04\hsize}X}
\newcolumntype{k}{>{\hsize=.96\hsize}X}
\newcolumntype{m}{>{\hsize=.2\hsize}X}
\newcolumntype{y}{>{\hsize=.33\hsize}X}
\begin{table*}[!ht]
    \centering
    \caption{Highlights of Findings from Survey Questions by Gender}
    \begin{tabularx}{\textwidth}{v|k}
        \hline
        RQ1 & What are your expectations from your organization?\newline Career Opportunities = CP, Goal Achievement = GA, Knack in Standard SE Practice = SP, Learning Opportunities = LO, Proper Management = PM, Work Benefits = WB, Work-place Culture = WC\newline
        1) \textbf{Male}: CO (20.51\%), GA (10.26\%), SP (10.26\%), LO (7.69\%), PM (30.77\%), WB (28.21\%), WC (48.72\%), 2) \textbf{Female}: CO (50.0\%), GA (0.0\%), SP (50.0\%), LO (0.0\%), PM (50.0\%), WB (50.0\%), WC (50.0\%)
        % {
        % \begin{tabularx}{0.92\textwidth}{yyyyyyyy}
        %     & CO (\%) & GA (\%) & SP (\%) & LO (\%) & PM (\%) & WB (\%) & WC (\%)\\ 
        % Male & 20.51 & 10.26 & 10.26 & 7.69 & 30.77 & 28.21 & 48.21 \\
        % Female & 50.0 & 0.0 & 50.0 & 0.0 & 50.0 & 50.0 & 50.0 \\
        % \end{tabularx}
        % }
        \\ \hline
        
        RQ2 & What are your expectations from your manager?\newline Career Opportunities = CO, People Skill = PS, Proper Management = PM, Tech Savvy = TS\newline
        1) \textbf{Male}: CO (19.79\%), PS (50.0\%), PM (43.75\%), TS (12.5\%), 2) \textbf{Female}: CO (33.33\%), PS (33.33\%), PM (44.44\%), TS (22.22\%),
        % {
        % \begin{tabularx}{0.92\textwidth}{yyyyy}
        %     & CO (\%) & PS (\%) & PM (\%) & TS (\%) \\
        % Female & 33.33 & 33.33 & 44.44 & 22.22 \\
        % Male & 19.79 & 50.0 & 43.75 & 12.5 \\
        % \end{tabularx}
        % }
        \\ \hline
        
        RQ3 & What are your expectations from peers in the team? \newline Knack in Standard SE Practice = SP, Learning Environment = LE, Sincerity = Sincerity, Supportive Attitude = SA\newline
        1) \textbf{Male}: SP (55.7\%), LE (10.13\%), Sincerity (44.3\%), SA (55.7\%), 2) \textbf{Female}: SP (33.33\%), LE (16.67\%), Sincerity (33.33\%), SA (50.0\%)
        % {
        % \begin{tabularx}{0.92\textwidth}{yyyyy}
        %     & SP (\%) & LE (\%) & Sincerity (\%) & SA (\%)\\
        % Female & 33.33 & 16.67 & 33.33 & 50.0 \\
        % Male & 55.7 & 10.13 & 44.3 & 55.7 \\
        % \end{tabularx}
        % }
        \\ \hline
        
        RQ4 & What are your expectations from the new hires?\newline
        Knack in Standard SE Practice = SP, Motivation for Learning = ML, Sincerity = Sincerity\newline
        1) \textbf{Male}: SP (50.0\%), ML (50.0\%), Sincerity (50.0\%), 2) \textbf{Female}: SP (0.0\%), ML (100.0\%), Sincerity (100.0\%)
        % {
        % \begin{tabularx}{0.92\textwidth}{yyyy}
        %     & SP (\%) & ML (\%) & Sincerity (\%)\\
        % Male & 50.0 & 50.0 & 50.0 \\
        % Female & 0.0 & 100.0 & 100.0 \\
        % \end{tabularx}
        % }
        \\ \hline
        
        RQ5 & What are your expectations from the universities and government?\newline
        Universities: Career Support = CS, Industry Oriented Teaching = IOT, Learning Environment = LE \newline
        1) \textbf{Male}: CS (25.76\%), IOT (78.79\%), LE (21.21\%), 2) \textbf{Female}: CS (50.0\%), IOT (50.0\%), LE (25.0\%)
        % {
        % \begin{tabularx}{0.92\textwidth}{yyyy}
        %     & CS (\%) & IOT (\%) & LE (\%) \\
        % Male & 25.76 & 78.79 & 21.21 \\
        % Female & 50.0 & 50.0 & 25.0 \\
        % \end{tabularx}
        % }
        %\\ \hline
        \newline
        %RQ5 & What are your expectations from the Government?\newline
        Government: Career Opportunities = CO, Policy and Facility = PF, Proper Education = PE, Support for Industry = SI\newline
        1) \textbf{Male:} CO (22.95\%), PF (62.3\%), PE (22.95\%), SI(31.15\%)
        2) \textbf{Female:} CO (50\%), PF (75\%), PE (25\%), SI (0\%)
         %{
         %\begin{tabularx}{0.92\textwidth}{yyyyy}
         %         & CO (\%) & PF (\%) & PE (\%) & SI (\%) \\ 
         %Male & 22.95 & 62.3 & 22.95 & 31.15 \\
         %Female & 50.0 & 75.0 & 25.0 & 0.0 \\
        %
         %\end{tabularx}
         %}
        \\ \hline
        
    \end{tabularx}
    \label{table:analysis_by_gender}
\end{table*}
\section{Discussions}
In this section, we first analyze our study findings by two demographics of survey participants: experience and gender (\sec\ref{sec:analysis_exp_gender}). We then discuss the implications of our study findings in \sec\ref{sec:implications}.
\subsection{Analysis by Experience and Gender}\label{sec:analysis_exp_gender}
% \gias{please see my paper to find how you can enhance this section (e.g., Table 12)}
In Table~\ref{table:analysis_by_experience_part1} we summarize the results by the reported experiences of survey participants. Around 51.5\% respondents have 5 or less than 5 years of experience, and 48.4\% respondents have more than 5 years of experience. We noticed that young employees (0-5 years of experience) expect \emph{knack in standard SE practice} more than senior employees. The observation may indicate a change in SE practice in the industry. However, the observation is not statistically significant (based on the Mann-Whitney U test $p=0.08$). Similarly, we have observed that young respondents are more concerned about their work benefits (based on the Mann-Whitney U test $p=<0.01$). Moreover, in the expectation from peers, we observed that senior employees expect more \emph{supportive attitude} (based on Mann-Whitney U test $p=0.004$) and \emph{sincerity} (based on Mann-Whitney U test $p<0.01$) than young employees.
% \subsection{Analysis by Gender}\gias{please add more information. is there any difference between the males and females in terms of the observed categories per RQ?}\partha{Added}

In Table~\ref{table:analysis_by_gender} we summarize the results of questions from our survey by the reported gender of survey participants. In the expectation from the organization, we have observed significant differences in `career opportunities’ and `knack in Standard SE practice' categories. We observed that female respondents expected more `career opportunities’ than male respondents in expectations from managers (Mann-Whitney U test $p=0.17$). Female respondents emphasized more on standard SE practices than the male respondents (Mann-Whitney U test $p=0.054$). However, the observations are not statistically significant. We observed that female respondents expected more `career opportunities’ than male respondents from their managers (based on Mann-Whitney U test $p=0.17$). The observation may indicate that female software engineers struggle in the SE industry. However, the observation is not statistically significant. Similarly, in the expectation from peers we observed that male practitioners expect relatively more sincerity from their peers (based on Mann-Whitney U test $p=0.3$). In the expectation from the government, we noticed female respondents expect more `career opportunities’ than male respondents (Mann-Whitney U test $p=0.11$). One of the reasons for the insignificant observation is the low number of female SE practitioners in the industry. 
% The obse supports our hypothesis that there is a subtle gender bias in the industry.
% Similar to Hussein et al.~\cite{Hussain2020} we can sense gender bias in the industry. However, the observations are not significant enough to draw any conclusion.

\subsection{Implication of Findings}\label{sec:implications}
The findings from our study can be useful to guide \begin{inparaenum}
\item \bf{Software organizations} to prioritize measures on expectation management of developers, 
\item \bf{Software Managers} to better communicate with the fellow developers to improve overall team cohesion and relationships with developers,
\item \bf{Software Developers} to understand how they can be better team player by meeting the expectations of peers, 
\item \bf{Government and educational institutions} to focus on specific developer-centric policies, facilities and educational curriculum, and 
\item \bf{SE researchers} to investigate new tools and techniques for developers productivity and well-being.
\end{inparaenum}

\begin{figure}[h]
\centering
\vspace{-5mm}
\includegraphics[scale=0.7]{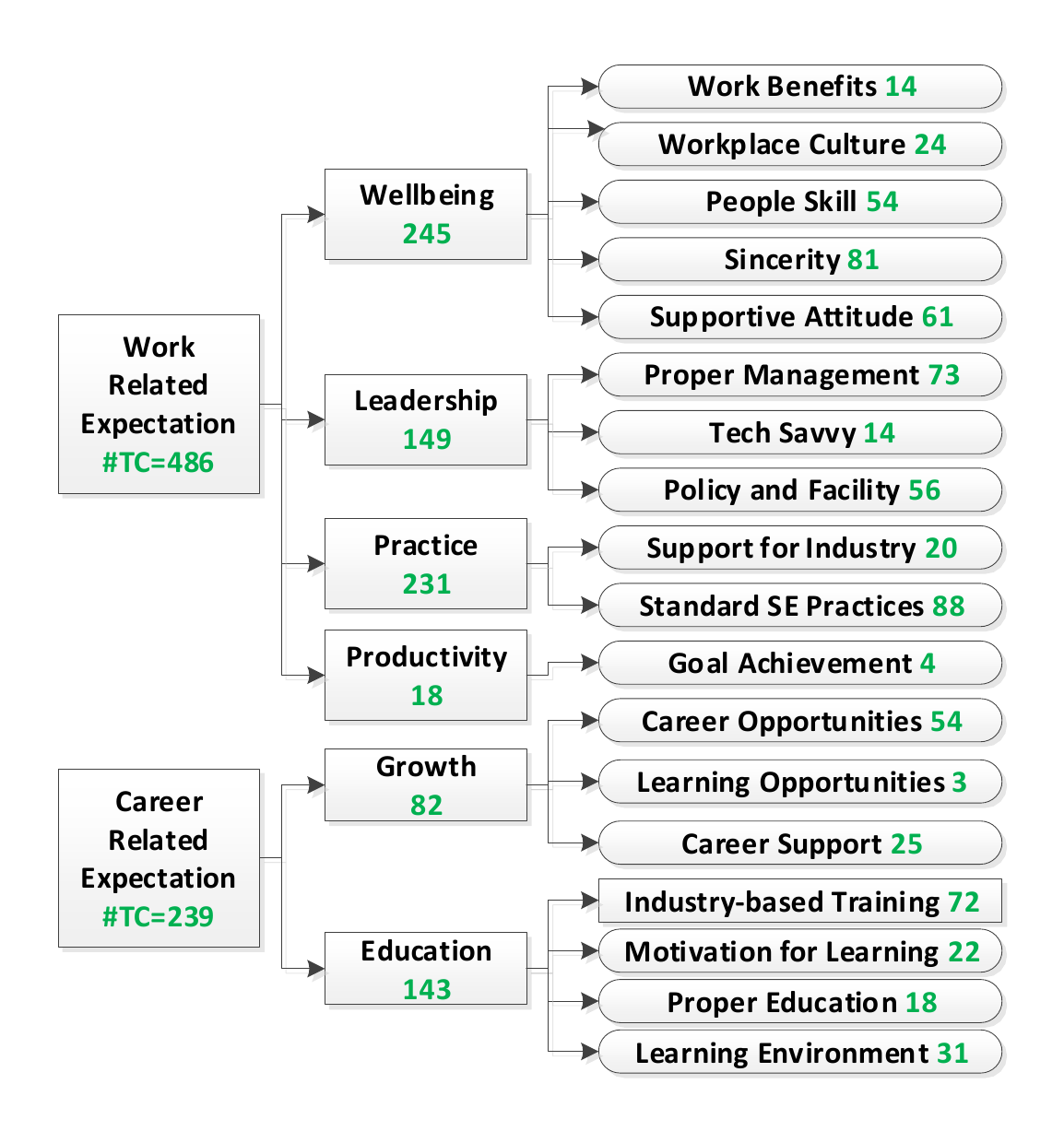}
\vspace{-5mm}
\caption{A hierarchy of developer expectations types as observed in the qualitative analysis of our survey responses}
\label{fig:taxonomy_TM}
\vspace{-3mm}
\end{figure}

\bf{\ul{Software Organizations.}} In \fig\ref{fig:taxonomy_TM}, we present a hierarchical view of 18 expectation types we observed in our study. We group the 18 expectation types in two categories: Work-related expectation and Career-related expectation. The work-related expectations are found in four themes: \begin{inparaenum}
\item Well-being (4 expectation types),
\item Leadership (3 types),
\item Practice (2 types),  and
\item Productivity (1 type).
\end{inparaenum} The career-related expectations belong to two themes: 
\begin{inparaenum}
\item Growth (3 types), and 
\item Education (3 types).
\end{inparaenum} Overall, we find more quotes for work-related expectations than for career-related expectations (486 vs 239). This observation shows that happiness and expectation management of developers cannot be met by simply offering them more money or promotion~\cite{Graziotin-HappinessProductivitySoftwareEngineer-BookRethinkingProductivity2019}. 
In \tbl\ref{table:key_findings}, we summarize the 18 expectation types by each RQ. We sort the expectation types (in descending order). For example, `Work-place culture' is mentioned by the most of the participants as an expectation type from their organization. Therefore, organization can learn from this study that good workplace culture ensuring work-life balance and alleviating the influence of corporate politics are the most desired characteristic among the developers, even more than the attraction of compensation package. Hence, they have to take special care to create and maintain such an environment. 
% It may be deduced that if the management decides to use the money saved from the expenditure of reasonably reduced compensation to create a better working environment, it would positively impact employee retention.
\begin{table}
  \centering
  \caption{Highlights of findings from our survey}
    \begin{tabular}{p{8cm}} \toprule
    \textbf{RQ1. Expectation from organization} \\ 
    1) Work-place Culture 48.8\%, 2) Proper Management 31.7\%, 3) Work Benefits 29.3\%, 4) Career Opportunities 22\%, 5) Standard SE Practice 12.2\%, 6) Goal Achievement 9.8\%, 7) Learning Opportunities 7.3\% \\
    \midrule 
    \textbf{RQ2. Expectation from manager} \\
    1) People Skill 48.6\%, 2) Proper Management 43.8\%, 3) Career Opportunities 21\%, 4) Tech Savvy 13.3\% \\
    \midrule
    \textbf{RQ3. Expectation from peers} \\
    1) Supportive Attitude 55.3\%, 2) Knack in Standard SE Practice 54.1\%, 3) Sincerity 43.5\%, 4) Learning Environment 10.6\% \\
    \midrule 
    \textbf{RQ4. Expectation from the new hires} \\
    1) Motivation for Learning 52.9\%, 2) Knack in Standard SE Practice 47.1\%, 3) Sincerity 47.1\% \\
    \midrule
    \textbf{RQ5. Expectation from government \& universities} \\
    Universities: 1) Industry Oriented Teaching 77.1\%, 2) Career Support 27.1\%, 3) Learning Environment 21.4\% \\
    Government: 1) Policy and Facility 63.1\%, 2) Support for Industry 29.2\%, 3) Career Opportunities 24.6\%, 4) Proper Education 23.1\% \\
    \bottomrule
    \end{tabular}%
  \label{table:key_findings}
  \vspace{-3mm}
\end{table}%

\bf{\ul{Software Management.}} In \fig\ref{fig:taxonomy_TM}, we find that expectations related to the leadership are found in 149 quotes (third most quoted theme). From \tbl\ref{table:key_findings}, we  see that leadership and management needs encompass the team managers as well as the government (e.g., policy support). In particular, while it is not easy for software team managers to be hands-on in the day to day development jobs of developers, our survey results show that software managers should work towards become tech savvy. However, the most important thing for them to be mindful of the socio-technical and career needs of the developers. The owners must take the relevant feedback of the developers regarding their expectations about management into account. This study reveals that the most important expectation from management is that they have to be good team players. Then come good managerial skills, the ability to create career opportunities for engineers, and technical proficiency. 

\bf{\ul{Software Developers.}} Given software development is a team activity, the success of a good product development relies heavily on team cohesion and understanding. The expectation from developers by their peers is revealed in this study. We see that both from new recruits and regular developers, sincerity about responsibilities, knack and ability to learn new technologies, and attitude to follow standard software engineering practice are desired. From the fellow developers, the most desired attribute is the supportive mindset. 
% As an individual member of a software development team, everyone should keep these in mind right from the beginning of his/her career.

\bf{\ul{Government and educational institutions.}}  Among the most number of
responses, we find expectations from educational institutions to offer relevant
teaching and from governments to improve job stability, which indicate the
increasingly important roles of these organizations to help software developers.
This observation can be especially true during the COVID-19 pandemic, which is when our study was conducted. We find that SE practitioners require policy and regulatory support to ensure job environment, funding for universities to improve research and practice modern curriculum and recognize the knowledge-based SE industry. 

\bf{\ul{SE Researchers.}} Our study results offer complementary perspective to the large body of research in developer productivity and well-being~\cite{Meyer-CharacterizeSoftDevPerceptionOfProductivity-ESEM2017,Nguyen-AnalysisTrendsProductivityCost-PMSE2011,Albrecht-MeasureApplicatonDevelopmentProductivity-IBM1979,Meyer-SoftDevPerceptionProductivity-FSE2014,Paiva-FactorsProductivitySoftDev-CSSE2010,Ko-WhyWeShouldNotMeasureProductivity-BookRethinkingProductivity2019,Perry-PeopleOrganizationProcessImprovement-IEEESw1994,Baruch-SelfPerformanceDirectManager-ManagerialPsychology1996,Chong-InterruptionsOnSoftwareTeams-CSCW2006,Czerwinski-DiaryStudyTaskSwitchingInterruptions-CHI2006,Parnin-CueForResumingInterruptedProgrammingTasks-CHI2010,Meyer-SoftDevPerceptionProductivity-FSE2014} by offering a high-level view of the diverse types of expectations that developers can have towards five stakeholders. Such insights can help design new studies into developer productivity, e.g., does interruption from novice vs expert colleagues mean different while measuring productivity? or does productivity improve when manager is more tech-savvy? 
% Most of our survey participants are from small or medium organizations, which complement current research conducted in big organizations.

% \subsubsection{Implication to the Universities}
% In countries with a developing economy like Bangladesh, having a decent job is often challenging. Most companies cannot afford to give training opportunities to the recruits and want them ready for professional practices. Whereas globally computer science and engineering departments of universities focus on providing a basic foundation of a wide range of topics and encouraging research, many students want more focused training on industry-popular skills in developing countries. This is a challenge that universities have to deal with and balance their curriculum to accommodate such ever-changing demands.

% \subsubsection{Implication to the Government}
% The study offers excellent insight to the Government to support the SE industry. The Government of a developing country like Bangladesh may conceive that the SE practitioners require policy and regulatory support to ensure job environment, funding for universities to improve research and practice modern curriculum and recognize the knowledge-based SE industry. This insight is expected to help relevant departments of the Government make the best use of the public fund.

\section{Threats to Validity}
\label{validity}
%\rifat{I have slightly modified the threats to validity}.
% We discuss the threats to validity of our studies following common guidelines
% for empirical studies~\cite{Woh00}.
\bf{Construct validity} is mainly concerned with the extent to which the study objectives truly represent the theory behind the study \cite{Wohlin2012}. In our study, we have used open coding strategy to label the survey responses. The nature of this coding strategy may introduce researcher bias into coded labels. To mitigate the issue, the labels have been coded by two individuals, and the codes are accepted when there is a reasonable agreement among the coders. It was previously observed\cite{Garousi2015} that people tend to form their answers close to expected answers when evaluated. To mitigate the threat, before the survey, we informed participants that our motive in this survey was to get a decent understanding of current practices, and we do not intend to collect any personally identifiable data. Construct threats may also be introduced by a misleading interpretation of the survey questions. We conducted a preliminary survey and interview session with some participants to rule out any ambiguity from survey questions and thus reduce such risk.
\bf{Internal validity} is a property of scientific studies that refers to how well a study has been conducted. A threat to internal validity in this study is inherent in the participant selection bias. We used personal connections to reach as many participants as possible. Another threat could arise from the placement of the options in a multiple-choice question. It is often observed that survey participants often show bias towards the first option in any multiple-choice question\cite{Uddin-SurveyOpinion-TSE2019}. However, in our case, all the multiple-choice questions were asked about the role and experience of the participants, and there was no concern of bias there. Moreover, from the personal practical experience of the authors, there is no bias in this opinion.
\bf{External validity} is concerned with the generalization of the study result. While more responses would have offered more proof of generalizability, we note that we already observed saturation in our manual coding of themes and labels (i.e., during open coding). We also found a considerable concentration of professionals supporting each aspect of expectations we studied in the paper. 

% However, it is difficult to claim the statistical generalizability of our findings, given that our sample included 137 respondents where there are 1100+ companies and 3,00,000 IT professionals\cite{BASIS2018} in the software industry of Bangladesh. In general, emerging IT industries share a common trend of challenges\cite{Sison2006, lloyd2020}. Thus, our findings are also applicable to other emerging software industries across the globe. \anindya{The rational behind writing the last two lines is not very clear.}
% \rifat{Gias vai, please have a look.}
\section{Conclusions}
In this paper, we surveyed 181 professional software developers to understand their expectations from five different stakeholders: \begin{inparaenum} 
\item organizations, 
\item managers, 
\item peers, 
\item new hires, and 
\item government and educational institutions.
\end{inparaenum} The five stakeholders are determined by conducting semi-formal interviews of software developers. We ask open-ended survey questions and analyze the responses using  open-coding. We observed 18 multi-faceted expectations types. While some expectations are more specific to a stakeholder, other expectations are cross-cutting. For example, developers  expect work-benefits from their organizations, but expect the adoption of standard software engineering (SE) practices from their organizations, peers, and new hires. Among the most number of responses, we find expectations from educational institutions to offer relevant teaching and from governments to improve job stability, which indicate the increasingly important roles of these organizations to help software developers. This observation can be especially true during the COVID-19 pandemic. Our future work will revisit developers' productivity and well-being measures to determine how the various expectation types we observed in our study could be used to improve/complement the measures.

\balance

\bibliographystyle{ACM-Reference-Format}
\bibliography{references}

\end{document}